%% file: Cheb-2011.TEX
\documentclass [winedt,yap]{iitparc}
\usepackage{cite}
\usepackage{amsmath,amssymb,amsfonts,bm}
\usepackage{eufrak}
\usepackage{lscape}
\usepackage{floatfig,wrapfig,epsfig}
\usepackage{subfigure}
\usepackage{color}
\usepackage{psboxit}
\usepackage{rotating}
\usepackage{curves}
\usepackage{ulem}
\usepackage[mathscr]{eucal}
\RequirePackage{psfrag}
\RequirePackage{graphicx}

\usepackage{amsxtra}
\usepackage{array}

\begin{document}\normalem
\initfloatingfigs
\frontmatter          

\IssuePrice{25.00}%
\TransYearOfIssue{2011}%
\TransCopyrightYear{2011}%
\OrigYearOfIssue{2011}%
\OrigCopyrightYear{2011}%

\TransVolumeNo{72}%
\TransIssueNo{12}%
\OrigIssueNo{12}%


\mainmatter

\setcounter{page}{2458}
\CRubrika{MULTI-AGENT SYSTEMS}
\Rubrika{MULTI-AGENT SYSTEMS}

%
\include{AgaChe}

\end{document}

%% file: AgaChe.tex

\title{The Projection Method for Reaching Consensus and\\
       the Regularized Power Limit of a Stochastic Matrix\thanks{This work was supported in part by the Russian
Foundation for Basic Research, project no.~09-07-00371 and the program of RAS Presidium ``Mathematical Theory of Control.''}}

\author{R. P. Agaev and P. Yu. Chebotarev}

\institute{Trapeznikov Institute of Control Sciences, Russian Academy of Sciences, Moscow, Russia}

\titlerunning{THE PROJECTION METHOD FOR REACHING CONSENSUS}

\authorrunning{Agaev, Chebotarev}

\OrigCopyrightedAuthors{R.P. Agaev and P.Yu. Chebotarev}

\received{Received February 22, 2011}

\OrigPages{pp.~38-59}

\maketitle

\def\rank{\mathop{{\rm rank}}\nolimits}          
\def\R{\mathbb{R}}                               
\def\C{\mathbb{C}}                               
\def\ind{\mathop{\rm ind}\nolimits}              
\def\NN{\mathop{\mathcal N}\nolimits}            
\def\RR{\mathop{\mathcal R}\nolimits}            
\def\sqa{\sqcap\!\!\!\!\sqcup}                   
\def\epr{\hfill$\sqa$\smallskip}                 
\def\l{\ell}                                     
\def\Pbes{P^\infty}                              
\def\aa{\alpha}                                  
\def\e{\varepsilon}                              
\def\G{\Gamma}                                   
\def\bb{\beta}                                   
\def\si{\sigma}                                  
\def\cdc{,\ldots,}                               
\def\1n{1,\ldots,n}                              
\def\J{\tilde{J}}                                
\def\ms{\mathstrut}                              
\def\xz{\hspace{-.07em}}                         
\def\xy{\hspace{.07em}}                          
\def\tr{\operatorname{tr}}                       
\accentedsymbol{\Pin}{\stackrel{\;\scriptscriptstyle\infty}{P_{\mathstrut}}} 
\def\Up#1{\vspace{-#1em}}                        
\def\beq{\begin{equation}}                       
\def\eeq{\end{equation}}                         
\def\O{0}                                        
\def\B{{\protect\mathfrak B}}                    

\newlength{\widebarargwidth}
\newlength{\widebarwidth}
\newlength{\widebarargheight}
\newlength{\widebarargdepth}
\DeclareRobustCommand{\wbar}[1]{%
  \settowidth{\widebarargwidth}{\ensuremath{#1}}%
  \settoheight{\widebarargheight}{\ensuremath{#1}}%
  \settodepth{\widebarargdepth}{\ensuremath{#1}}%
  \addtolength{\widebarargwidth}{-0.7\widebarargheight}
  \addtolength{\widebarargwidth}{-3.8\widebarargdepth}
  \makebox[0pt][l]{\addtolength{\widebarargheight}{-0.2ex}
    \hspace{0.2\widebarargheight}%
    \hspace{0.2\widebarargdepth}%
    \hspace{0.5\widebarargwidth}%
    \setlength{\widebarwidth}{0.6\widebarargwidth}%
    \addtolength{\widebarwidth}{0.6ex}
    \makebox[0pt][c]{\rule[\widebarargheight]{\widebarwidth}{0.1ex}}}
  {#1}}

\begin{abstract}
In the coordination/consensus problem for multi-agent systems, a well-known condition of achieving consensus is the presence of a spanning arborescence in the communication digraph. The paper deals with the discrete consensus problem in the case where this condition is not satisfied. A~characterization of the subspace $T_P$ of initial opinions (where $P$ is the influence matrix) that \emph{ensure\/} consensus in the DeGroot model is given. We propose a method of coordination that consists of: (1)~the transformation of the vector of initial opinions into a vector belonging to $T_P$ by orthogonal projection and (2)~subsequent iterations of the transformation~$P.$ The properties of this method are studied.
It is shown that for any non-periodic stochastic matrix $P,$ the resulting matrix of the orthogonal projection method can be treated as a regularized power limit of~$P.$
\end{abstract}

\section{INTRODUCTION}

In the last decade, the problem of reaching consensus in multi-agent systems has been the subject of many works.
For the basic results in the field, we refer to the surveys and monographs \cite{RenBeardAtkins07,Olfati-SaberFaxMurray07,CheAga09ARC,AgaChe10UBSE,Wu10IEEE,GuNoCh10E,Wu07book,Jackson08,RenBeard08bookCopy,MesbahiEgerstedt10book,RenCao11distributed}.

One of the first discrete models of reaching consensus was proposed by DeGroot~\cite{DeGroot74}. Suppose that $s(0)=(s_{1}^0\cdc s_{n}^0)^{\rm T}$ is the vector of initial opinions of the members of a group and ${s(k)=(s_{1}^k\cdc s_{n}^k)^{\rm T}}$ is the vector of their opinions after the $k$th step of coordination. In accordance with the DeGroot model, $s(k)=P s(k-1),\;k=1,2,\ldots,$ where $P$ is a row stochastic influence matrix whose entry $p_{ij}$ specifies the degree of influence of agent $j$ on the opinion\footnote{Thus, row $i$ of $P$ determines an iterative adjustment process for agent $i$'s opinion; since $P$ is row stochastic, the degrees of influence on each agent sum to~1.} of agent~$i.$
Thereby,
\beq\label{281110eq1}
s(k)=P^k s(0),\quad k=1,2,\ldots.
\eeq

Consensus is [asymptotically] \emph{achieved} if $\lim_{k\to\infty}s_i^k=\bar s$ for some $\bar s\in\R$ and all $i\in\{\1n\}.$ It has been shown \cite{DeGroot74} that consensus is achieved for any initial opinions if and only if the matrix $\Pbes=\lim_{k\to\infty}P^k$ exists and all rows of $\Pbes$ are identical, which is equivalent to the regularity\footnote{A stochastic matrix is said to be \emph{regular\/}~\cite{Gantmacher60} if it has no eigenvalues of modulus~1 except for the simple eigenvalue~$1.$ Regular stochastic matrices are also called \emph{SIA\/} (stochastic, indecomposable, aperiodic). In the terminology of matrix analysis, we mainly follow~\cite{Gantmacher60,HornJohnson86}; in the terminology of graph theory we follow \cite{Harary69,Tutte84,CvetkovicDoobSachs80}.} of~$P$. Thus, reaching consensus in the DeGroot model is determined by the asymptotic properties of the powers of~$P$ and the initial opinions.

If $P$ is not regular, then the opinions do not generally tend to agreement. Yet, consensus can be achieved if the vector of initial opinions belongs to a certain subspace. Below we characterize this subspace and consider the \emph{method of projection\/} which ensures that consensus is achieved even if the vector of initial opinions does not belong to the above-mentioned subspace. Furthermore, it is found that the resulting matrix of the orthogonal projection procedure can be treated as a regularized power limit of the initial stochastic matrix.

The paper is organized as follows.
After introducing the terminology (Section\:\ref{s_Notat}) and listing a number of well-known results used in the analysis of network dynamics (Section\:\ref{s_prelim}), in Section\:\ref{s_Deso} we discuss the conditions of reaching consensus in the DeGroot model. Section\:\ref{s_chardo} characterizes the region of convergence to consensus for the DeGroot model.
Sections\:\ref{s_metho}--\ref{s_nebaz} present the orthogonal projection procedure which generalizes the DeGroot algorithm and can be applied when this algorithm does not ensure that consensus is achieved. These sections also give the structure of the projection and the properties of the proposed method.
In Section\:\ref{s_neorto}, we consider some non-orthogonal projections onto the subspace of convergence to consensus. In Section\:\ref{s_nebaz}, we demonstrate that in the orthogonal projection procedure (as well as in the DeGroot algorithm), the nonbasic agents do not affect the final result. In sections\:\ref{s_disj} and\:\ref{s_neba}, the case of the absence of nonbasic agents is studied and the main result is extended to the general case. Section\:\ref{s_interp} briefly discusses the interpretation of the orthogonal projection procedure, and the final Section\:\ref{s_regu} deals with the concept of the regularized power limit of a stochastic matrix.

\section{BASIC NOTATION}
\label{s_Notat}

With a stochastic influence matrix $P,$ we associate the \emph{communication digraph\/} $\G$ with vertex set $V(\G)=\{\1n\}.$ $\G$~has the $(j,i)$ arc with weight $w_{\xz ji}=p_{ij}$ whenever $p_{ij}>0$ (i.e., whenever agent $j$ influences agent~$i$). Thus, arcs in $\G$ are oriented \emph{in the direction of influence}; the weight of an arc is the power of influence.

The \emph{Kirchhoff matrix\/} (see \cite{Tutte84,CheAga02ap}) $L=L(\G)=(\l_{ij})$ of digraph $\G$ is defined as follows: if $j\ne i,\,$ then $\l_{ij}=-w_{\xz ji}$ whenever $\G$ has the $(j,i)$ arc and $\l_{ij}=0$ otherwise; $\l_{ii}=-\sum_{k\ne i}\l_{ik},\,$ $i={\1n}$. The Kirchhoff matrix has zero row sums and nonpositive off-diagonal entries.

The matrices of this kind are sometimes referred to as \emph{directed Laplacians\/}~\cite{VeermanFlocks06}. However, in a more precise terminology~\cite[Section~2.2]{CheAga02ap}, the Laplacian matrix of a digraph is the matrix with zero row sums whose non-diagonal entries are defined by $\l_{ij}=-w_{ij},$ i.e., as distinct from the Kirchhoff matrix, the entries in the $i$th row are determined by the weights of the arcs \emph{outgoing\/} from\footnote{Because of the similarity of these definitions, the Kirchhoff and Laplacian matrices are often confused. In the problems of decentralized control, either formalism can be used. If the analysis is based on the construction of the influence digraph (as in the present paper), it is convenient to use Kirchhoff matrices, while if the digraph of references (requests for information) is constructed whose arcs are directed oppositely, it is more convenient to use Laplacian matrices.}~$i.$
Consequently, if all arcs of $\G$ are reversed, then the Laplacian matrix of the resulting digraph coincides with the Kirchhoff matrix of $\G$  and vice versa. Consequently, the Kirchhoff matrices and Laplacian matrices of digraphs form the same class.

By virtue of the above definitions, for the digraph $\G$ associated with $P$ we have
\beq\label{e_L:I-P}
L(\G)=I-P,
\eeq
where $I$ is the identity matrix.

Any maximal by inclusion strong 
(i.e., with mutually reachable vertices) subgraph of a digraph is called a {\it strong component} (or a {\it bicomponent}) of this digraph. A {\it basic bicomponent\/} is a bicomponent such that the digraph has no arcs coming into this bicomponent from outside. Vertices belonging and not belonging to basic bicomponents can be called {\it basic\/} and {\it nonbasic\/}, respectively. Similarly, we call an agent {\it basic}/{\it nonbasic} when the vertex representing this agent is {\it basic}/{\it nonbasic}. Let $b$ and $\nu$ be the number of basic vertices and the number of basic bicomponents in $\G,$ respectively.

We enumerate the basic bicomponents and after them the nonbasic bicomponents. Furthermore, we enumerate the vertices in the first bicomponent, next the vertices in the second bicomponent, and so on. We shall assume that agents are numbered correspondingly. In this case, the influence matrix $P$ and the Kirchhoff matrix $L$ have a lower block-triangular form called the \emph{Frobenius normal form}. In the matrices $P$ and $L$ represented in this form, the upper-left blocks of size $b\times b$ correspond to the basic vertices of the communication digraph. These blocks will be denoted by $P_\B$ and $L_\B$:
\beq
\label{051110eqa}
P=\left(\begin{array}{cc}
  P_\B & 0  \\
   \ast&\ast\\
\end{array}\right),\quad
L=\left(\begin{array}{cc}
  L_\B & 0  \\
   \ast&\ast\\
  \end{array}\right).
\eeq
$P_\B$ and $L_\B$ respectively coincide with the influence matrix and the Kirchhoff matrix of the communication digraph restricted to the set of basic vertices/agents.

The vertex set of any bicomponents is called a \emph{class}. We also speak of the corresponding \emph{classes of agents}. The vertex set of a basic bicomponent will be called a \emph{final class\/}\footnote{In the above context, this definition may seem illogical, since a basic bicomponent does not allow arcs from outside. However, it is justified by the fact that in the Markov chain determined by the influence matrix $P,$ transitions occur not in the direction of the arcs of influence, but rather in the direction of agents that influence. Thereby, ``all roads lead to'' the basic bicomponents, and the union of final classes is exactly the set of essential states of the corresponding Markov chain.}~\cite{BermanShakedMonderer09}. 

If the DeGroot algorithm converges and vertex $j$ is not basic, then, as noted in \cite{DeGroot74}, column $j$ in the limiting matrix $\Pbes$ is zero and the initial opinion of agent $j$ does not affect the limiting vector of opinions.

\section{USEFUL RESULTS FROM ALGEBRAIC GRAPH THEORY}
\label{s_prelim}

In this section, we present a number of results that are useful in the analysis of the DeGroot model and other network control models. In particular, they are used to prove the subsequent theorems and propositions (the proofs are given in the Appendix).

First, if the sequence of powers $P^k$ of a stochastic matrix $P$ has a limit $\Pbes,$ then
\beq\label{e_Pbes_L}
\Pbes=\J,
\eeq
where $\J$ is the normalized matrix of maximum out-forests of the corresponding weighted digraph~$\G$ (a corollary of the matrix tree theorem for Markov chains~\cite{WentzellFreidlin70a}).

The matrix $\J$ is equal to the matrix $\J_{n-\nu}$ defined recursively:
\beq\label{e_Jk}
\J_k=I-k\frac{L\J_{k-1}}{\tr(L\J_{k-1})},\;\;\text{where}\;\;k=\1n-\nu,\;\;\J_0=I,\;\;\text{and}\;\;L\J_{n-\nu}=0
\eeq
(see \cite[Section\:4]{AgaChe01} or \cite[Section\:5]{CheAga02ap})
and can also be found by passage to the limit:
\beq\label{e_Jli}
\J=\lim_{\tau\to\infty}(I+\tau L)^{-1}
\eeq
(Theorem~6 in \cite{AgaChe00}).

Furthermore, since $L=I-P,$ we have
\beq\label{e_PLLP}
\Pbes L= L\Pbes=0
\eeq
and
\beq\label{e_compl}
\NN(\Pbes)=\RR(L),\quad \RR(\Pbes)=\NN(L),
\eeq
where $\NN(A)$ and $\RR(A)$ are the kernel (null space) and the range of $A$, respectively (see, e.g., \cite[Section~5]{AgaChe01}).
Moreover, $\Pbes$ is the eigenprojection\footnote{On the methods of computing eigenprojections, see, e.g., \cite{AgaChe02,CheAga11ARC}.}
(principal idempotent) of $L$~\cite{Meyer75,Rothblum76ai} and
\beq\label{e_ranks}
\rank\Pbes=\nu;\quad\rank L=n-\nu,
\eeq
where $\nu$ is the number of basic bicomponents in~$\G$ \cite[Proposition~11]{AgaChe00}. It follows from \eqref{e_ranks} that
\beq\label{e_dimNL}
\dim\NN(L)=\nu,
\eeq
where $\dim\NN(L)$ is the dimension of the kernel (the \emph{nullity}) of~$L.$ Finally, by \cite[Proposition~12]{CheAga02ap},
\beq\label{e_NLRL}
\NN(L)\cap\RR(L)=\{\bm0\},
\eeq
\beq\label{e_indL}
\ind L=1,
\eeq
where $\ind L$ (the index of $L$) is the order of the largest Jordan block of $L$ corresponding to the zero eigenvalue, and by \eqref{e_dimNL} and~\eqref{e_indL},
\beq\label{e_mult0}
m_L(0)=\nu,
\eeq
where $m_L(0)$ is the multiplicity of $0$ as an eigenvalue of~$L$.

\section{CONDITIONS FOR CONVERGENCE TO CONSENSUS IN THE DEGROOT MODEL}
\label{s_Deso}

As noted in the Introduction, the DeGroot algorithm converges to consensus for any initial opinions if and only if there exists a limiting matrix ${\Pbes=\lim_{k\to\infty}P^k}$ having all rows equal. According to the ergodic theorem for Markov chains, a necessary and sufficient condition for this is, in turn, the regularity (the SIA property) of~$P.$

The equality of all rows of $\Pbes$ amounts to
\beq
\label{e_Pbes1}
\Pbes=\bm1\pi^{\rm T}
\eeq
with some probability vector (the components are non-negative and sum to~$1$)~$\pi,$ where
$\bm1=(1\cdc 1)^{\rm T}.$ In this case, the consensus $\bar s$ is expressed by the inner product of the vectors $\pi$ and~$s(0)$:
\beq
\label{e_pis}
s(\infty)=\Pbes s(0)=\bm1\pi^{\rm T}s(0)=\bm1\bar s,
\eeq
where $s(\infty)$ is the limiting vector of opinions, $\pi$ is the limiting weight distribution of the DeGroot algorithm, and ${\bar s=\pi^{\rm T}s(0)}$ is the consensus.

A probability vector $\pi$ is called a {\it stationary vector} of a stochastic matrix $P$ if it is a left eigenvector of $P$ corresponding to the eigenvalue\:$1$: $\pi^{\rm T}P=\pi^{\rm T}$.
Obviously, this condition is satisfied for the vector $\pi$ in the representation $\Pbes=\bm1\pi^{\rm T}$ of $\Pbes,$ provided that the convergence of the DeGroot algorithm to consensus is guaranteed by the regularity of~$P.$

By Theorem~3 in \cite{DeGroot74}, if for any vector of initial opinions $s(0),$ the DeGroot algorithm converges to the consensus $\pi^{\rm T}s(0),$ then\footnote{In fact, already in \cite{Gantmacher60} (\S\,7 of Chapter\:13) it was observed that if $P$ is regular, then the vector $\pi$ can be uniquely recovered from the equation $\pi=P^{\rm T}\pi$ and each row of $\Pbes=\lim_{k\to\infty}P^k$ is the transpose of~$\pi.$}
$\pi$ is a \emph{unique\/} stationary vector of~$P$.

Let us mention some \emph{sufficient\/} conditions \cite{DeGroot74} of the convergence of $P^k$ ($k\to\infty$) to a matrix with identical rows. One of them is the presence of an entirely positive column (stochastic matrices with at least one positive column are called \emph{Markov matrices}) in $P^k$ for some~$k.$ Another sufficient condition is that all states in the Markov chain corresponding to $P$ are mutually accessible  (in this case, $\G$ is strongly connected and the agents belong to the same class) and $P$ is \emph{proper\/} \cite{Gantmacher60} (which means that $P$ has no eigenvalues of modulus $1$ that are not equal to~$1$); in this case, $P$ is said to be \emph{primitive}.

A criterion of convergence to consensus for the DeGroot algorithm can also be formulated in terms of the communication digraph~$\G.$
The equality of the rows of $\Pbes$ is equivalent to $\,\rank\Pbes=1.$ Therefore, owing to \eqref{e_ranks}, when the sequence $\{P^k\}$ converges,  consensus is achieved for any initial opinions if and only if the communication digraph $\G$ corresponding to $P$ has a single basic bicomponent ($\nu=1$). Consequently, provided that the sequence $P^k$ converges, $\nu=1$ is equivalent to the regularity of~$P$. In turn, by \eqref{e_mult0}, this is the case if and only if $0$ is a simple eigenvalue of~$L.$

Finally, $\nu=1$ if and only if $\G$ has a spanning \emph{out-tree} (also called \emph{arborescence\/} and \emph{branching}) \cite[Proposition\:6]{AgaChe00}.
In this case (see\,\eqref{e_Pbes_L}), $\Pbes\!=\!(p_{\ms ij}^{\ms\scriptscriptstyle\infty})\!=\!\J\!=\!(\J_{ij})$ is the normalized matrix of spanning out-trees~\cite{AgaChe00}:
\beq
\label{e_J1}
p_{ij}^{\ms\scriptscriptstyle\infty}=\pi_j=\J_{ij}=\frac{t_j}t,\quad i,j=\1n,
\eeq
where $t_j$ is the total weight\footnote{The weight of an out-tree (and, more generally, of a digraph) is the product of the weights of all its arcs.} of $\G$'s spanning out-trees rooted at $j\,$ and $t$ is the total weight of all spanning out-trees of~$\G.$ It follows from \eqref{e_pis} and \eqref{e_J1} that in the case of guaranteed consensus,
\begin{gather*}
\bar s=\left(\frac{t_1}t\cdc \frac{t_n}t\right)s(0)=t^{-1}\sum_{j=1}^nt_j\,s_j^0.
\end{gather*}

A survey of some results on the DeGroot model and its generalizations can be found in~\cite{Jackson08,AgaChe10UBSE}. Note that one of the applications of the DeGroot model is information control in social networks~\cite{BarabanovKorNovChk10}.

\section{THE REGION OF CONVERGENCE TO CONSENSUS OF THE DEGROOT ALGORITHM}
\label{s_chardo}

Consider an influence matrix $P$ whose powers converge to\footnote{This assumption is not too restrictive.
It is satisfied, e.g., if every final class has at least one agent taking into account its own current opinion in the iterative adjustment of its opinion.
More generally, the limit $\Pbes$ exists~\cite{Gantmacher60} if and only if $P$ is \emph{proper}, i.e., $P$ has no eigenvalues ($\ne1$) of modulus~$1.$} $\Pbes,$ but the rows of $\Pbes$ are not necessarily equal.

In the vector space of initial opinions $s(0),$  let us find the subspace $T_P$ whose vectors are transformed by $\Pbes$ into vectors with equal components. Obviously, the DeGroot algorithm \eqref{281110eq1} with a proper matrix $P$ leads to a consensus if and only if $s(0)\in T_P.$ That is why $T_P$ will be referred to as the \emph{region of convergence to consensus of the DeGroot algorithm\/}~\eqref{281110eq1}. In Section\:\ref{s_metho}, we will present a consensus procedure which consists of two steps: on the first step, the vector of initial opinions $s(0)\not\in T_P$ is replaced by the nearest vector in~$T_P$; on the second step, the algorithm~\eqref{281110eq1} is applied to the result of the first step.

The following theorem characterizes the subspace~$T_P$.

\begin{theorem}
\label{240810pr1}
If the powers $P^k$ of the stochastic matrix $P$ converge\/$,$ then $T_P=\RR(L)\oplus T_1,$ where $T_P$ is the region of convergence to consensus of the DeGroot algorithm~\eqref{281110eq1}$,$ $L=I-P,\,$ and $T_1$ is the linear span of the vector $\bm1=(1\cdc 1)^{\rm T}.$
\end{theorem}

Now we reformulate Theorem\:\ref{240810pr1} in a different form.

\begin{corollary}\label{c_Li}
Suppose that the powers $P^k$ of the stochastic matrix $P$ converge and ${L=I-P.}$ Let $L^{(i)}$ and $M^{(i)}_\xi$ be the matrices resulting from $L$ by substituting $\bm1$ for the $i$th column and adding $\xi\bm1,$ where $\xi\in\R\smallsetminus\{0\},$ to the $i$th column\/$,$ respectively. Then $\,T_P=\RR(L^{(i)})=\RR\!\big(M^{(i)}_\xi\big)$ for any $i=\1n$.
\end{corollary}

To prove Corollary\:\ref{c_Li}, it is sufficient observe that any column of $L$ is equal to the sum of the remaining columns taken with the minus sign. Consequently, its removal does not affect the linear span of the columns. Thus, $\RR(L^{(i)})=\RR(L)\oplus T_1$ and so ${T_P=\RR(L^{(i)})}$ is equivalent to the assertion of Theorem\:\ref{240810pr1}. Similarly, $\RR\!\big(M^{(i)}_\xi\big)=\RR(L)\oplus T_1,$ and $T_P=\RR\!\big(M^{(i)}_\xi\big)$ is also equivalent to the assertion of Theorem\:\ref{240810pr1}.

\begin{remark}
\label{r_sisu}
Corollary\:\ref{c_Li} can also be formulated as follows: $s(0)\in T_P$ if and only if either system of equations $L^{(i)}x=s(0)$ or $M^{(i)}_\xi x=s(0)$ is consistent. Indeed, the consistency of these systems is tantamount to $s(0)\in\RR(L^{(i)})$ and $s(0)\in\RR\!\big(M^{(i)}_\xi\big),$ respectively.
\end{remark}

In what follows, we will need a matrix with independent columns whose range is~$T_P.$

\begin{corollary}\label{c_Li1}
Let $U$ be any matrix obtained from $L$ by\,
$(1)$~deleting\/$,$ for each final class\/$,$ one column corresponding to some vertex of this class and\,
$(2)$~adding $\bm1$ as the first column. Then the columns of\/ $U$ are independent and $\,T_P=\RR(U).$ 
\end{corollary}

\begin{remark}
\label{r_neba}
Owing to Corollary\:\ref{c_Li1}, the $\,n-b$ columns of $L$ corresponding to the nonbasic vertices are linearly independent. Therefore, for $U$ to remain a matrix of full column rank, they can be replaced by any $n-b$ independent columns with zeros in all ``basic'' rows (see also Proposition\:\ref{p_TP} and Corollary\:\ref{c_Desca} below).
\end{remark}

Using \eqref{e_compl} and \eqref{e_Pbes_L} the region $T_P$ of convergence to consensus can also be represented through the kernel of ${\Pbes=\J.}$

\begin{corollary}\label{c_J}
Under the assumptions of Theorem~$\ref{240810pr1},\,$ $T_P=\NN(\J)\oplus T_1.$
\end{corollary}

\begin{remark}
\label{r_00}
According to~\eqref{e_ranks} $\rank\J=\nu.$ By virtue of \cite[(102) in Chapter\:13]{Gantmacher60}, $\Pbes=\J$ has a lower block-triangular form. Moreover, its submatrix corresponding to the basic vertices is block-diagonal: every basic bicomponent is represented by a diagonal block with equal rows. Therefore, $\NN(\J)$ is the orthogonal complement of the linear span of $\nu$ columns of $\J^{\rm T}$ taken one from each diagonal block representing a basic bicomponent. 
\end{remark}

The following proposition enables one to ``locate'' the subspace~$T_P.$

\begin{proposition}
\label{p_TP}
Let $x\in\R^n.$ Then $x\in T_P\Leftrightarrow x^{}_\B\in T_{P_\B},$ where $x^{}_\B$ is $x$ with the nonbasic components removed.
\end{proposition}

By the definition of $T_P,$ to prove Proposition\:\ref{p_TP}, it is sufficient establish that
$\Pbes x=a\bm1_n\Leftrightarrow(P_\B)^\infty x^{}_\B=a\bm1_b,$ where $a\in\R,$ $(P_\B)^\infty=\lim_{k\to\infty}(P_\B)^k,$ and the vectors $\bm1$ are supplied with their dimensions as a subscript. This equivalence can be easily deduced from the following properties of the matrix $\Pbes$ (see \cite[(102) in Chapter\:13]{Gantmacher60}): (1)~the submatrix of $\Pbes$ corresponding to the basic vertices coincides with $(P_\B)^\infty$; (2)~the last $n-b$ columns of $\Pbes$ consist of zeros; (3)~the last $n-b$ rows of $\Pbes$ are convex combinations of the first $b$ rows.

\begin{corollary}[of Proposition\:\ref{p_TP}]
\label{c_Desca}
The region of convergence to consensus of the DeGroot algorithm has the form $\,T_P\!=\!T_{P_\B}\!\times\R^{n-b}.$
\end{corollary}

\begin{example}\label{Po_zakazu}{\rm
Consider the multi-agent system whose communication digraph $\G$ is shown in Fig.\,\ref{f_syst1}.

\begin{figure}[t]
\centering{\includegraphics[height=1.825in]{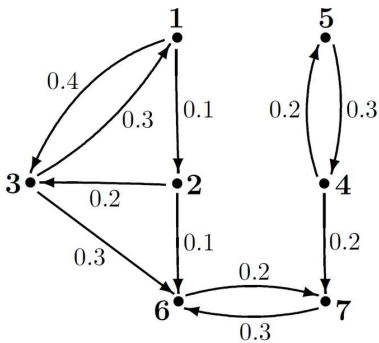}}
\caption{The communication digraph $\G$ of a multi-agent system.\label{f_syst1}}
\end{figure}
For simplicity, Fig.\,1 does not show loops (every vertex has a loop whose weight sums to $1$ with the weights of all arcs \emph{entering\/} this vertex).

The basic bicomponents of $\G$ are the restrictions of $\G$ to the classes $\{1,2,3\}$ and $\{4,5\}$; the class $\{6,7\}$ is nonbasic.
Matrices $P$ and $L=I-P$ of the communication digraph $\G$ are as follows:
{\small
\[
P=\left(\begin{array}{>{\!}rrrrrrr}
0{.}7&0    &0{.}3&0    &0    &0    &0    \\
0{.}1&0{.}9&0    &0    &0    &0    &0    \\
0{.}4&0{.}2&0{.}4&0    &0    &0    &0    \\
0    &0    &0    &0{.}7&0{.}3&0    &0    \\
0    &0    &0    &0{.}2&0{.}8&0    &0    \\
0    &0{.}1&0{.}3&0    &0    &0{.}3&0{.}3\\
0    &0    &0    &0{.}2&0    &0{.}2&0{.}6\\
\end{array}\!\right),\quad
L=\left(\!\!\!\begin{array}{>{\!}rrrrrrr}
 0{.}3& 0    &-0{.}3& 0    & 0    & 0    & 0    \\
-0{.}1& 0{.}1& 0    & 0    & 0    & 0    & 0    \\
-0{.}4&-0{.}2& 0{.}6& 0    & 0    & 0    & 0    \\
 0    & 0    & 0    & 0{.}3&-0{.}3& 0    & 0    \\
 0    & 0    & 0    &-0{.}2& 0{.}2& 0    & 0    \\
 0    &-0{.}1&-0{.}3& 0    & 0    & 0{.}7&-0{.}3\\
 0    & 0    & 0    &-0{.}2& 0    &-0{.}2& 0{.}4\\
\end{array}\!\right).
\]
}

We construct the matrix $U$ (see Corollary\:\ref{c_Li1}) by removing the first and the fourth column of $L$ and adding $\bm1$ as the first column; also find the matrix $\Pbes=\J$ by using \eqref{e_Jk} or \eqref{e_Jli} or by computing $\lim_{k\to\infty}P^k$\,:
{\small
\beq
\label{e_UJ}
U\!=\!\left(\begin{array}{>{\!}r>{\!}r>{\!}r>{\!}r>{\!}r>{\!}r>{\!}r}
1& 0    &-0{.}3& 0    & 0    & 0    \\
1& 0{.}1& 0    & 0    & 0    & 0    \\
1&-0{.}2& 0{.}6& 0    & 0    & 0    \\
1& 0    & 0    &-0{.}3& 0    & 0    \\
1& 0    & 0    & 0{.}2& 0    & 0    \\
1&-0{.}1&-0{.}3& 0    & 0{.}7&-0{.}3\\
1& 0    & 0    & 0    &-0{.}2& 0{.}4\\
\end{array}\!\right),\quad
\Pbes\!\approx\!\left(\!\begin{array}{rrrrrrr}
0{.}4  &0{.}4  &0{.}2  &0      &0      &0&0\\
0{.}4  &0{.}4  &0{.}2  &0      &0      &0&0\\
0{.}4  &0{.}4  &0{.}2  &0      &0      &0&0\\
0      &0      &0      &0{.}4  &0{.}6  &0&0\\
0      &0      &0      &0{.}4  &0{.}6  &0&0\\
0{.}291&0{.}291&0{.}146&0{.}109&0{.}164&0&0\\
0{.}146&0{.}146&0{.}073&0{.}255&0{.}382&0&0\\
\end{array}\!\right)
\eeq
}
(in $\Pbes,$ all entries, except for the decimal values in the last two rows, are exact).

According to Corollary\:\ref{c_Li1}, $\xy U$ has full column rank and the linear span of its columns coincides with the region of convergence to consensus of the DeGroot algorithm: $\RR(U)=T_P$. Finally, according to Corollary\:\ref{c_J} and Remark\:\ref{r_00}, $\,T_P$ is the direct sum of $T_1$ and the orthogonal complement of the linear span of the vectors\footnote{Here, the row vectors are represented in a ``matrix'' form, i.e., their components are separated by a space.}
\beq
\label{e_p1p2}
\tilde\pi^1=\bigl(0{.}4\;\;0{.}4\;\;0{.}2\;\;0    \;\;0    \;\;0\;\;0\bigr)^{\rm T}\;\;\text{and}\;\;
\tilde\pi^2=\bigl(0    \;\;0    \;\;0    \;\;0{.}4\;\;0{.}6\;\;0\;\;0\bigr)^{\rm T},
\eeq
obtained by transposing the rows of~$\Pbes$ corresponding to the different final classes.
} 
\end{example}

\section{THE ORTHOGONAL PROJECTION METHOD}
\label{s_metho}

If $P$ is not regular, then $T_P\ne\R^n,$ i.e., there are vectors of initial opinions not driven to consensus by the DeGroot algorithm.
Consider the case where consensus is still needed. How can it be reached?
A natural suggestion is to enrich the communication digraph with additional links that will ensure regularity of the matrix $P$ and to perform the  iterative adjustment of opinions with the new matrix. However, one can imagine a situation where communication between agents is their prerogative and the only thing the ``center'' may suggest (or theoretically consider) is a certain correction of the initial conditions~$s(0).$ In what follows, we consider mathematical and algorithmic aspects of this approach.

As shown above, to ensure reaching consensus by means of the DeGroot algorithm with matrix $P,$ it is necessary to transform the vector $s(0)$ into some vector $s'(0)\in T_P.$ In doing so, it is desirable to minimize $\|s'(0)-s(0)\|,$ where $\|\!\cdot\!\|$ is the Euclidean norm.

The transformation mapping any vector $s(0)$ into the closest vector in $T_P$ is the orthogonal projection of $\R^n$ onto $T_P$ (along the orthogonal subspace $T_P^\bot$). According to Lemma~2.3 in \cite{Ben-IsraelGreville7403} this projection is given by a symmetric idempotent matrix; we will denote it by~$S$.

If the initial conditions are adjusted by means of the orthogonal projection $S,$ then the limiting vector of opinions $s(\infty)$ can be represented as follows:\footnote{Note that in finding $s(\infty)$, as well as in the DeGroot algorithm, there is no necessity to compute the powers of $P$: it is sufficient to iterate the multiplication of $P$ by \emph{vectors\/}, starting with the vector $Ss(0)$; the calculation of $Ss(0)$ will be discussed below.}
\begin{gather*}
s(\infty)=\Pbes Ss(0).
\end{gather*}

This method will be called the \emph{orthogonal projection procedure\/} for reaching consensus.
The matrix $\Pbes S$ will be referred to as the \emph{resulting matrix of the orthogonal projection procedure\/} and denoted by~$\Pin\!:$
\begin{gather}
\label{e_Pin}
\Pin=\Pbes S.
\end{gather}

By construction, the orthogonal projection procedure leads to a consensus for any vector of initial opinions%
\footnote{Formally, this can be proved as follows: by the definition of $S,$ Theorem\:\ref{240810pr1}, and Eq.\,\eqref{e_PLLP},\,
$\RR(\Pbes S)=\{\Pbes Ss(0)\mid s(0)\in\R^n\}
 =\{\Pbes(Ly+a\bm1)\mid y\in\R^n,a\in\R\}
 =\{a\bm1\mid a\in\R\}.$}
\!$s(0)$ (recall that $T_P$ is the region of convergence to consensus of the DeGroot algorithm, while $S$ projects $\R^n$ onto~$T_P$). Consequently, all rows of the matrix $\Pin$ are identical, i.e., there is a vector $\aa=(\aa_1\cdc\aa_n)^{\rm T}$ such that
\beq
\label{e_Pinaa}
\Pin=\bm1\aa^{\rm T}.
\eeq
Vector $\aa$ will be referred to as the \emph{weight vector of the orthogonal projection procedure}.
Thus, the proposed procedure can be written in the form
\beq
\label{e_aas}
s(\infty)=\Pbes Ss(0)=\Pin s(0)=\bm1\aa^{\rm T}s(0)=\bm1\bar s,
\eeq
where $s(0)$ is an arbitrary vector of initial opinions, $\aa$ is the weight vector of the orthogonal projection procedure, and $\bar s=\aa^{\rm T}s(0)$ is the consensus.

Is $\aa$ a probability vector? $S$~is the projection onto the subspace $T_P$ which by Theorem\:\ref{240810pr1} contains~$\bm1.$ Consequently, $S$ leaves $\bm1$ fixed, i.e., $S$ has all row sums equal to~$1.$ Given the fact that $\Pbes$ is stochastic, we obtain that ${\Pin=\Pbes S}$ also has row sums~$1,$ i.e., $\sum_{i=1}^n\aa_i=1.$ Observe, however, that $S$ may have negative entries. Therefore, the answer to the question on the nonnegativity of the weight vector $\aa$ is not immediately obvious.
This question will be answered in Section\:\ref{s_disj}. This will enable one to interpret the matrix~$\Pin$ as the regularized power limit of the stochastic matrix~$P.$

\section{FINDING THE ORTHOGONAL PROJECTION}
\label{s_mapro}

Consider the properties of the projection~$S.$
It is known \cite{Ben-IsraelGreville7403} that for any rectangular matrix $A,$ the matrix $AA^+$ where $A^+$ is the Moore-Penrose generalized inverse of $A,$ is the orthogonal projection with range~$\RR(A)$.

Note that the matrix $U$ defined in Corollary\:\ref{c_Li1} has full column rank ${n-\nu+1}$ and $\RR(U)=T_P$.
Hence, owing to the above fact, the orthogonal projection $S$ with range $T_P$ has the expression $S=UU^+$. To determine $U^+,$ we use the formula $U^+=(U^{\rm T} U)^{+}U^{\rm T}$ (see, e.g., Problem~2.17(d) in \cite{Ben-IsraelGreville7403}) which for the matrix $U$ of full column rank takes the form $U^+=(U^{\rm T} U)^{-1}U^{\rm T}$. Consequently,
\beq
\label{17102010eq1}
S=UU^+=U(U^{\rm T} U)^{-1}U^{\rm T}.
\eeq

The following proposition clarifies the structure of~$S.$
\begin{proposition}
\label{p_STp}
The orthogonal projection $S$ onto the subspace $T_P$ has the form
\beq
\label{e_Sbl}
S=\left(\begin{array}{cc}S_\B&\O\\
                                    \O&I\\\end{array}\right),
\eeq
where $S_\B$ is the orthogonal projection onto the region $T_{P_\B}$ of convergence to consensus of the DeGroot algorithm with matrix~$P_\B.$
\end{proposition}

Proposition\:\ref{p_STp} follows from Corollary\:\ref{c_Desca}. The nonbasic components of any vector do not alter when the vector is projected onto $T_P,$ since any alteration would be contrary to the minimality of the distance to the projection.

Computing, with the help of \eqref{17102010eq1}, the projection $S$\/ for Example\:\ref{Po_zakazu} we obtain
{\small
\[
S=
\frac1{22}\left(\begin{array}{rrrrrrr}
18&-4&-2& 4& 6& 0&0\\
-4&18&-2& 4& 6& 0&0\\
-2&-2&21& 2& 3& 0&0\\
 4& 4& 2&18&-6& 0&0\\
 6& 6& 3&-6&13& 0&0\\
 0& 0& 0& 0& 0&22&0\\
 0& 0& 0& 0& 0&0&22\\
\end{array}\right)
\approx\left(\!\begin{array}{rrrrrrr}
 0{.}818&-0{.}182&-0{.}091& 0{.}182& 0{.}273&0&0\\
-0{.}182& 0{.}818&-0{.}091& 0{.}182& 0{.}273&0&0\\
-0{.}091&-0{.}091& 0{.}955& 0{.}091& 0{.}136&0&0\\
 0{.}182& 0{.}182& 0{.}091& 0{.}818&-0{.}273&0&0\\
 0{.}273& 0{.}273& 0{.}136&-0{.}273& 0{.}591&0&0\\
 0      & 0      & 0      & 0      & 0      &1&0\\
 0      & 0      & 0      & 0      & 0      &0&1\\
  \end{array}\!\right)\!.
\]
} 

\noindent Substituting $S$ and the matrix $\Pbes=\J$ \eqref{e_UJ} into \eqref{e_Pin} we find the resulting matrix of the orthogonal projection procedure:
\beq
\label{e_Pinex}
\Pin=\Pbes S
=          \bm1\!\cdot\!\tfrac1{110}\bigl(26      \;\;      26\;\;      13\;\;      18\;\;      27\;\;0\;\;0\bigr)
\!\approx\!\bm1\!\cdot\!            \bigl(0{.}2364\;\;0{.}2364\;\;0{.}1182\;\;0{.}1636\;\;0{.}2455\;\;0\;\;0\bigr).\!\!\!
\eeq

Note the following properties of the vector $\aa=\tfrac1{110}\bigl(26\;26\;13\;18\;27\;0\;0\bigr)^{\rm T}$ representing the matrix $\Pin$ in accordance with~\eqref{e_Pinaa}:
(1)~the components of $\aa$ corresponding to the basic vertices are strictly positive; the components corresponding to the nonbasic vertices are zero; (2)~$\sum_{i=1}^n\aa_i=1$;
(3)~as the comparison of \eqref{e_Pinex} and \eqref{e_p1p2} suggests, if vertices $k$ and $m$ belong to the $i$th final class, then $\aa_k/\aa_m=\tilde\pi^i_k/\tilde\pi^i_m,$ where $\tilde\pi^i$ is the stationary vector of the influence matrix of the $i$th basic bicomponent.

The second property has already been proved in the general case (Section\:\ref{s_metho}). Later in this paper we will prove that the remaining properties hold true as well.

Generally speaking, the orthogonal projection procedure makes a transition from the matrix $\Pbes$ bringing each final class to its separate consensus to the matrix $\Pin$ establishing a common consensus.

\section{DO THE NONBASIC AGENTS AFFECT ANYTHING?}
\label{s_nebaz}

As noted in the Introduction, if the DeGroot algorithm leads to a consensus, then this consensus does not depend on the initial opinions of the nonbasic agents.
This property is inherited by the orthogonal projection procedure with the difference that under this procedure with a proper (aperiodic) matrix $P,$ consensus is always achieved. This is stated by the following proposition.

\begin{proposition}
\label{p_nearbird}
In the case of a proper matrix $P,$ the resulting vector of opinions $s(\infty)$ of the orthogonal projection procedure does not depend on the initial opinions of the nonbasic agents.
\end{proposition}

Note that according to~\eqref{e_aas}, $s(\infty)=\Pbes s'(0),$ where $s'(0)=S s(0).$ By Proposition\:\ref{p_STp}, the components of $s'(0)$ corresponding to the nonbasic vertices are equal to the corresponding components of~$s(0)$: the \emph{``preequalization''} performed by the transformation $S$ does not alter the opinions of the nonbasic agents.

Proposition\:\ref{p_nearbird} immediately follows from the equation $s(\infty)=\Pin s(0)$ \eqref{e_aas} and the following representation of the matrix~$\Pin.$

\begin{theorem}
\label{p_Pin}
For any proper matrix $\,P,$ vector $\,\aa$ determining the resulting matrix\/ $\Pin=\bm1\aa^{\rm T}$ of the orthogonal projection procedure has the form
\begin{gather*}
\aa=(\aa_1\cdc\aa_b,0\cdc0)^{\rm T},
\end{gather*}
\noindent where $(\aa_1\cdc\aa_b)$ is any row of the matrix $(P_\B)^\infty S_\B$ $($see \eqref{051110eqa} and \eqref{e_Sbl}$).$
\end{theorem}

The answer to the more general question ``Is the result of the orthogonal projection procedure affected by the \emph{presence\/} of nonbasic agents?'' is also ``No.''

\begin{proposition}
\label{p_nearbirdd}
In the case of a proper matrix $P,$ consensus $\bar s$ of the orthogonal projection procedure does not alter with the exclusion of nonbasic agents\/$,$ provided that the initial opinions of the basic agents\/$,$ as well as the weights of their influence on each other$,$ are preserved.
\end{proposition}

Proposition\:\ref{p_nearbirdd}, as well as Proposition\:\ref{p_nearbird}, follows from Theorem\:\ref{p_Pin} which implies that the vector $(\aa_1\cdc\aa_b)^{\rm T}$ determining $\Pin$ does not alter with the exclusion of nonbasic agents. The properties of this vector will be summarized in Theorem\:\ref{250810th1} (Section\:\ref{s_disj}).

Thus, under the orthogonal projection procedure, the only result of the presence of nonbasic agents is that finally their opinions come to the same consensus as the opinions of the basic ones. This consensus does not depend on the initial opinions of the nonbasic agents or even their presence.

\section{NONORTHOGONAL PROJECTION ON THE SUBSPACE OF CONVERGENCE TO CONSENSUS}
\label{s_neorto}

In the orthogonal projection procedure, iterative adjustment \eqref{281110eq1} is preceded by the projection of the vector of initial opinions $s(0)$ onto the subspace~$T_P$ (preequalization). 
By virtue of Proposition\:\ref{p_STp}, to find the orthogonal projection matrix $S,$ the communication digraph can be restricted to the set of basic vertices.

An alternative to $S$ is a stochastic matrix that transforms any vector of initial opinions into a vector in $T_P$ and at the same time approximates~$P.$ It makes sense to additionally require that this matrix be idempotent, since otherwise it would alter some vectors already in $T_P$ which do not need any preequalization. 
The problem of finding such a matrix has much in common with the classical problem of matrix approximation (see, e.g., \cite{EckartYoung36} and \cite[Section~7.4]{HornJohnson86}).

It can be shown that instead of $S$ one can use the matrix $\widetilde S$ that has the first $b$ rows equal to the corresponding rows of $S$ and the remaining rows equal to the last $n-b\,$ rows of~$P$. More specifically, if
$
P=\left(\begin{array}{cc}P_\B&\O\\
                                     B&D\\\end{array}\right)\;\;\text{and}\;\;
S=\left(\begin{array}{cc}S_\B&\O\\
                         \O&I\\\end{array}\right)
$
(see\:\eqref{051110eqa} and \eqref{e_Sbl}), then

\beq
\label{e_Sp}
\widetilde S=\left(\begin{array}{cc}S_\B&\O\\
                                                B&D\\\end{array}\right).
\eeq

$\widetilde S$ is not generally idempotent, however, since $B\ne \O,$ we have
$$\|P-\widetilde S\|_E
=\left\|\begin{array}{cc}P_\B-S_\B&\O  \\
                                                  \O&\O  \\\end{array}\right\|_E
<\left\|\begin{array}{cc}P_\B-S_\B&\O  \\
                           B&D-I\\\end{array}\right\|_E
=\|P-S\|_E,$$
where $\|X\|_E$ is the Euclidean norm of~$X,$ i.e., $\widetilde S$ is closer to $P$ than~$S.$
On the other hand, since $\Pbes$ has a zero block corresponding to the nonbasic vertices (indeed, they correspond to the inessential states of the Markov chain determined by~$P$; see\:\cite[(102) in Chapter\:13]{Gantmacher60}), we have $\Pbes S=\Pbes\widetilde S,$
consequently, the following proposition holds.

\begin{proposition}
\label{p_Sp}
If $P$ is proper and $\widetilde S$ is given by \eqref{e_Sp}$,$ then the two-stage procedure of coordination consisting of the preequalization $s'(0)=\widetilde Ss(0)$ and the iterative adjustment $s(k)=P^ks'(0),\,$ ${k=1,2\cdc}$ leads to the same consensus as the orthogonal projection procedure.
\end{proposition}

As a consequence, we obtain that $\RR(\widetilde S)\subseteq T_P.$
It can be shown that if the nonsingularity of the block $D$ is satisfied, then $\RR(\widetilde S)=T_P.$

\begin{remark}
\label{r_joke}
Let us mention other specific transformations of the space of initial opinions $\R^n$ into the subspace~$T_P.$ They do not require computing the orthogonal projection~$S.$ Owing to Corollary\:\ref{c_Li} of Theorem\:\ref{240810pr1}, for the matrices $L^{(i)}$ and $M^{(i)}_\xi$ obtained from $L$
by substituting $\bm1$ for the $i$th column and adding to it $\xi\bm1,$ where $\xi\in\R\smallsetminus\{0\},$ respectively, $\RR(L^{(i)})=\RR(M^{(i)}_\xi)=T_P$ holds.
Thus, preequalization of the initial vector $s(0)$ by means of either $L^{(i)}$ or $M^{(i)}_\xi$ ensures achieving consensus in the subsequent iterative adjustment with~$P.$ However, it is easy to verify that this approach generates dictatorial procedures: they lead to a consensus that is \emph{equal\/} to the initial opinion of the $i$th agent if $L^{(i)}$ or $M^{(i)}_1$ is used or is \emph{proportional\/} to it (i.e., distorts it) in the case of $M^{(i)}_\xi$ with $\xi\not\in\{0,1\}.$
Thus, this method is only good for the concealment of the dictatorial ``coordination'' of opinions.
\end{remark}

In considering other possible mappings of the space $\R^n$ into~$T_P,$ the main advantages of the projection $S$ should be taken into account, namely, that it guarantees the minimal difference between the initial vector of opinions and the result of its preequalization.

The orthogonal projection $S$ has the expression~\eqref{17102010eq1}. In Section\:\ref{s_disj}, we will obtain another explicit expression for~$S$ which is useful for studying the properties of the orthogonal projection procedure and its interpretation.

\section{THE ORTHOGONAL PROJECTION METHOD WHEN ALL AGENTS ARE BASIC}
\label{s_disj}

According to Theorem\:\ref{p_Pin} nonbasic agents do not affect the vector $(\aa_1\cdc\aa_b)^{\rm T}$ which determines the consensus in the orthogonal projection procedure. Therefore, all essential problems can be seen in the case where all agents are basic. This case is considered in the present section.

Suppose, as before, that $P$ is proper, $\lim_{k\to\infty}P^k=\Pbes,$ and the number of basic bicomponents is~$\nu.$ Since there are no connections between the basic bicomponents, in the absence of nonbasic agents the DeGroot algorithm is divided into $\nu$ independent processes of coordination.
It is interesting to see how the orthogonal projection procedure integrates these processes. 

Let $m_i$ be the number of vertices in the $i$th bicomponent. Its influence matrix, Kirchhoff matrix, and the power limit of the influence matrix we denote by $P_i,$ $L_i=(\l^i_{uv}),$ and $\Pbes_i,$  respectively.
In the absence of nonbasic agents, $P,$ $L,$ and $\Pbes$ have the form:
\beq
\label{191010eq1}
P=\left(\!\!\begin{array}{c>{\!}c>{\!}c>{\!}c}
P_1&   &      &     \\
   &P_2&      &     \\
   &   &\ddots&     \\
   &   &      &P_\nu\\
\end{array}\!\!\right)\!,\,\,\,
L=\left(\!\!\begin{array}{c>{\!}c>{\!}c>{\!}c}
L_1&   &      &     \\
   &L_2&      &     \\
   &   &\ddots&     \\
   &   &      &L_\nu\\
\end{array}\!\!\right)\!,\,\,\,
\Pbes=\left(\!\!\begin{array}{c>{\!}c>{\!}c>{\!}c}
\Pbes_1&       &      &         \\
       &\Pbes_2&      &         \\
       &       &\ddots&         \\
       &       &      &\Pbes_\nu\\
\end{array}\!\!\right)\!,
\eeq
where the diagonal blocks correspond to the final classes and the entries not in these blocks are zero.

The matrices $\Pbes_i$ correspond to the strongly connected digraphs, so (see the sufficient conditions of achieving consensus in Section\:\ref{s_Deso}) all the rows of each of them are identical, i.e., these matrices can be represented in the form
\beq
\label{e_Pbesi}
\Pbes_i=\bm1(\pi^i)^{\rm T},\quad i=1\cdc\nu,
\eeq
where $(\pi^i)^{\rm T}$ is any row of~$\Pbes_i.$

Consider the following problem: How is the weight vector $\aa$ of the orthogonal projection procedure \eqref{e_aas} related to the vectors~$\pi^i$?

Let ${q^i}=\tilde\pi^{i-1}-\tilde\pi^i,$ $i=2\cdc\nu,$ where $\tilde\pi^i\in\R^n$ is $\pi^i$ supplemented by zeros in the positions corresponding to all bicomponents, except for the $i$th one. Define $X$ as the matrix obtained from $L$ by replacing the first column by $\bm1$ and the first columns of the blocks corresponding to the other bicomponents by the zero columns. Thus, $X$ contains $\nu-1$ zero columns and all other columns of $X$ are independent, so $\rank X=n-\nu+1$ (see Corollary\:\ref{c_Li1}\, to Theorem\:\ref{240810pr1}).

Now define $Z$ as the matrix obtained from $X$ by replacing the zero columns in the blocks with numbers ${i=2\cdc\nu}$ by the vectors~$q^i.$ The form of $Z$ with $\nu=3$ is presented in Eq.\,\eqref{e_Z}:
\beq
\label{e_Z}
Z=\left(\begin{array}{ccccccccccccccccc}
1     &\l^1_{12}   &\ldots&\l^1_{1m_1}  &\pi^1_1     & 0         &\dots & 0           & 0          & 0          &\dots & 0            \\
\bm1  &\vdots      &\ddots&\vdots       &\vdots      &\bm0       &\dots &\bm0         &\bm0        &\bm0        &\dots &\bm0          \\
1     &\l^1_{m_1 2}&\dots &\l^1_{m_1m_1}&\pi^1_{m_1} & 0         &\dots & 0           & 0          & 0          &\dots & 0            \\
1     & 0          &\dots & 0           &-\pi^2_1    &\l^2_{12}  &\ldots&\l^2_{1m_2}  &\pi^2_1     & 0          &\dots & 0            \\
\bm1  &\bm0        &\dots &\bm0         &\vdots      &\vdots     &\ddots&\vdots       &\vdots      &\bm0        &\dots &\bm0          \\
1     & 0          &\dots & 0           &-\pi^2_{m_2}&\l^2_{m_22}&\ldots&\l^2_{m_2m_2}&\pi^2_{m_2} & 0          &\dots & 0            \\
1     & 0          &\dots & 0           & 0          & 0         &\dots & 0           &-\pi^3_1    &\l^3_{1 2}  &\ldots&\l^3_{1 m_3}  \\
\bm1  &\bm0        &\dots &\bm0         &\bm0        &\bm0       &\dots &\bm0         &\vdots      &\vdots      &\ddots&\vdots        \\
1     & 0          &\dots & 0           & 0          & 0         &\dots & 0           &-\pi^2_{m_3}&\l^3_{m_3 2}&\ldots&\l^3_{m_3 m_3}\\
\end{array}\right).
\eeq

\smallskip 
\begin{lemma}\label{041110pr1}
$Z$ is nonsingular.
\end{lemma}

By virtue of Lemma\:\ref{041110pr1} and the proof of Theorem~2.8 in \cite{Ben-IsraelGreville7403}, the orthogonal projection $S$ with range $T_P$ satisfies
\beq
\label{e_SXZ}
S=XZ^{-1}.
\eeq
Substituting \eqref{e_SXZ} into \eqref{e_Pin} and using \eqref{e_Pinaa} yield
\beq
\label{e_aaZ}
\bm1\aa^{\rm T}=\Pin=\Pbes S=\Pbes XZ^{-1}.
\eeq

Equation~\eqref{e_aaZ} is used in Theorem\:\ref{250810th1} which summarizes some properties of the vector $\aa,$ the weight vector of the orthogonal projection procedure (see~\eqref{e_aas}). In particular, this theorem establishes a relation between $\aa$ and the vectors~$\pi^i$ and matrix~$Z^{-1}$ (see\:\eqref{e_Pbesi} and \eqref{e_Z}).

\begin{theorem}
\label{250810th1}
If all agents are basic\/$,$ then the following assertions hold\/$:$

{\rm 1.} The row vector $\aa^{\rm T}$ coincides with the first row of $Z^{-1};$

{\rm 2.} All components of $\aa$ are positive and\, $\sum_{i=1}^n\aa_i=1.$ The sum of the entries in any row of $Z^{-1},$ except for the first row\/$,$ is\/~$0${\rm;}

{\rm 3.} Let $c(g)$ be the number of the bicomponent containing vertex~$g.$ Then for $g,h=\1n,$
$$
\frac{\aa_g}{\aa_h}=\frac{\bb_{c(g)}\,\tilde\pi_g^{c(g)}}{\bb_{c(h)}\,\tilde\pi_h^{c(h)}},
$$
where $\bb_i=(t^i)^2/\sum_{l=1}^{m_i}(t_l^i)^2${\rm;} $t^i$ and $t_l^i$ are the total weight of all spanning out-trees in the $i$th bicomponent of\/ $\G$ and the total weight of those of them that are rooted at~$l,$ respectively.
\end{theorem}

\begin{figure}[t]
\centering{\includegraphics[height=1.34in]{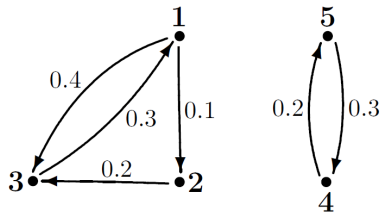}}
\caption{The communication digraph $\G_{\!\B}.$\label{f_syst1a}}
\end{figure}

\Up{1}
\begin{example}
\label{Po_zakazu2}
{\rm
Restrict the communication digraph of Example\:\ref{Po_zakazu} to the union of the final classes (Fig.\,\ref{f_syst1a}).
%
In other words, consider the subsystem whose influence matrix and Kirchhoff matrix have the form:
{\small
\[
P_\B
=\left(\begin{array}{rrrrr}
0{.}7&0    &0{.}3&0    &0    \\
0{.}1&0{.}9&0    &0    &0    \\
0{.}4&0{.}2&0{.}4&0    &0    \\
0    &0    &0    &0{.}7&0{.}3\\
0    &0    &0    &0{.}2&0{.}8\\
\end{array}\right),\quad
L_\B
=I-P_\B
=\left(\!\begin{array}{>{\!}r>{\!}r>{\!}r>{\!}r>{\!}r}
 0{.}3& 0    &-0{.}3& 0     & 0    \\
-0{.}1& 0{.}1& 0    & 0     & 0    \\
-0{.}4&-0{.}2& 0{.}6& 0     & 0    \\
 0    & 0    & 0    & 0{.}3 &-0{.}3\\
 0    & 0    & 0    &-0{.}2 & 0{.}2\\
  \end{array}\right).
\]
}

The limiting matrix $(P_\B)^\infty=\lim_{k\to\infty}(P_\B)^k$ of this subsystem is the ``basic'' submatrix of $\Pbes=\J$ which was found in Section\:\ref{s_chardo} (see\:\eqref{e_UJ}):
\[
(P_\B)^\infty
=\J_\B
=\left(\begin{array}{rrrrr}
0{.}4&0{.}4&0{.}2&0    &0    \\
0{.}4&0{.}4&0{.}2&0    &0    \\
0{.}4&0{.}4&0{.}2&0    &0    \\
0    &0    &0    &0{.}4&0{.}6\\
0    &0    &0    &0{.}4&0{.}6\\
  \end{array}\right).
\]

Find the resulting matrix of the orthogonal projection procedure using~\eqref{e_SXZ}. First, we construct the matrices $X$ and $Z$ defined above:
{\small
\begin{equation*}
X=\left(\begin{array}{rrrrr}
1& 0    &-0{.}3&0& 0    \\
1& 0{.}1& 0    &0& 0    \\
1&-0{.}2& 0{.}6&0& 0    \\
1& 0    & 0    &0&-0{.}3\\
1& 0    & 0    &0&0{.}2 \\
\end{array}\right),\quad
Z=\left(\begin{array}{rrrrr}
1& 0    &-0{.}3& 0{.}4& 0    \\
1& 0{.}1& 0    & 0{.}4& 0    \\
1&-0{.}2& 0{.}6& 0{.}2& 0    \\
1& 0    & 0    &-0{.}4&-0{.}3\\
1& 0    & 0    &-0{.}6& 0{.}2\\
\end{array}\right).
\end{equation*}
}

Compute $Z^{-1}$ and, using Eq.\,\eqref{e_SXZ}, the projection $S_\B$:
{\small
\[
Z^{-1}\!\approx\!\left(\begin{array}{>{\!\!}r>{\!}r>{\!}r>{\!}r>{\!}r}
 0{.}236&0{.}236& 0{.}118& 0{.}164& 0{.}245\\
-4{.}182&5{.}818&-2{.}091& 0{.}182& 0{.}273\\
-1{.}939&1{.}394& 0{.}697&-0{.}061&-0{.}091\\
 0{.}455&0{.}455& 0{.}227&-0{.}455&-0{.}682\\
 0{.}182&0{.}182& 0{.}091&-2{.}182& 1{.}727\\
\end{array}\!\!\right),\;
S_\B\!\approx\!
\left(\begin{array}{>{\!\!}r>{\!}r>{\!}r>{\!}r>{\!}r}
 0{.}818&-0{.}182&-0{.}091& 0{.}182& 0{.}273\\
-0{.}182& 0{.}818&-0{.}091& 0{.}182& 0{.}273\\
-0{.}091&-0{.}091& 0{.}955& 0{.}091& 0{.}136\\
 0{.}182& 0{.}182& 0{.}091& 0{.}818&-0{.}273\\
 0{.}273& 0{.}273& 0{.}136&-0{.}272& 0{.}591\\
 \end{array}\!\right)\!.
\]
}
$S_\B$ coincides (see Proposition\:\ref{p_STp}) with the ``basic'' submatrix of the projection $S$ found in Section\:\ref{s_mapro}.

The resulting matrix of the orthogonal projection procedure is given by~\eqref{e_Pin}:
\[
\Pin_\B=(P_\B)^\infty S_\B
\approx\bm1\!\cdot\!\bigl(0{.}2364\;\;0{.}2364\;\;0{.}1182\;\;0{.}1636\;\;0{.}2455\bigr).
\]
In accordance with Theorem\:\ref{p_Pin}, $\Pin_\B$ coincides with the ``basic'' submatrix of $\Pin$ (see\:\eqref{e_Pinex}).
The first row of $Z^{-1}$ coincides with any row of~$\Pin_\B,$ as stated in Theorem\:\ref{250810th1}.

Thus, in the case where all agents are basic, the orthogonal projection method makes a transition from a series of local consensuses reached by applying the transformation $(P_\B)^\infty$ to a global consensus established by~$\Pin_\B.$ A~relation between the vector $\aa$ determining $\Pin_\B$ with the vectors $\pi^i$ that determine $(P_\B)^\infty$ is given by item\:3 of Theorem\:\ref{250810th1}.
}
\end{example}

\section{ON THE INTERPRETATION OF THE ORTHOGONAL PROJECTION METHOD}
\label{s_interp}

An interpretation of the orthogonal projection method can be obtained using the following consequence of Theorem\:\ref{250810th1}: $\aa$ (the weight vector of the orthogonal projection procedure: $\Pin=\bm1\aa^{\rm T}$) is a probability vector; in other words, it determines an \emph{impact distribution\/} on the set of agents. 
Indeed, the resulting consensus is equal to the weighted mean of the initial opinions with weights taken from the distribution~$\aa.$

By Theorem\:\ref{250810th1}, the relative weight of the $i$th vertex of the $k$th bicomponent in the distribution $\aa$ is determined by (1)~the fraction of the weight of out-trees starting from vertex $i$ in the total weight of out-trees of the $k$th bicomponent (this fraction is equal to $\pi_k^i$) and (2)~the relative uniformity of the $k$th bicomponent w.r.t.\ the weight of the trees starting from its vertices: it can be shown that the higher the uniformity, the grater~$\bb_i.$ This leads to the conclusion that the resulting opinion is maximally influenced by those leaders in the bicomponents who ``broke away from their pursuers.''
Note that Eq.\,\eqref{e_aparo} gives several expressions for $\aa_g/\aa_h$ which provide a number of variations in interpretation of agents' weights in the orthogonal projection procedure.

In fact, the two-stage method consisting of the preequalization and the subsequent averaging using the influence matrix acts as if the basic bicomponents were combined into a single component by some additional arcs.
In a subsequent paper, we shall specify what additional arcs lead to a result equivalent to the preequalization using the orthogonal projection.

Note that enriching connections and finding the stationary vector of the stochastic matrix corresponding to the resulting digraph is the essence of the PageRank method~\cite{LangvilleMeyer06}. In it, the links are added evenly, between all vertices and with equal weights.
The structure of additional connections that emulates the orthogonal projection procedure is more complex.

\section{MORE ON NONBASIC AGENTS}
\label{s_neba}

The projection $S$ can be represented in the form \eqref{e_SXZ} not only in the case where all agents are basic, but in the general case  as well. For this, it is sufficient to apply the definitions of $X$ and $Z$ given in Section\:\ref{s_disj} to an arbitrary matrix~$L$ and use the proof of Theorem~2.8 in~\cite{Ben-IsraelGreville7403}. In this case, item\:1 of Theorem~\ref{250810th1} remains true.

Below we "obtain a somewhat more general result.
For the set of basic vertices, we define the matrix $X_\B$ using the definition of $X$ given in Section\:\ref{s_disj}. Now to define the matrix $X$ for the whole set of vertices, we supplement $X_\B$ with a zero block on the right, the block $L_{\wbar\B}$ that corresponds in $L$ to the nonbasic vertices, and an arbitrarily filled block\footnote{Both the block $G$ and the zero block on the right can simply be copied from the Kirchhoff matrix~$L.$}~$G$; let $Z$ be the same extension of~$Z_\B$:
\beq
\label{e_XZgen}
X=\left(\begin{array}{cc}X_\B&\O                   \\
                                      G&L_{\wbar\B}\\\end{array}\right),\quad
Z=\left(\begin{array}{cc}Z_\B&\O                   \\
                                      G&L_{\wbar\B}\\\end{array}\right).
\eeq

Due to Lemma\:\ref{041110pr1} in Section\:\ref{s_disj}, $Z_\B$ is nonsingular. According to the proof of Proposition~11 in \cite{AgaChe00}, $L_{\wbar\B}$ is also nonsingular. Therefore, $Z$ is invertible. Using \eqref{e_XZgen}, the Frobenius formula for the inversion of block matrices \cite[(86) in Chapter\:2]{Gantmacher60}, \eqref{e_SXZ}, and~\eqref{e_Sbl} we find
$$
XZ^{-1}
=\left(\begin{array}{cc}                   X_\B      &\O               \\
                                         G           &L_{\wbar\B}      \\\end{array}\right)\!
 \left(\begin{array}{cc}                   Z_\B^{-1} &\O               \\
                        -L_{\wbar\B}^{-1}G Z_\B^{-1} &L_{\wbar\B}^{-1} \\\end{array}\right)
=\left(\begin{array}{cc}              X_\B Z_\B^{-1} &\O               \\
                                                  \O &I                \\\end{array}\right)
=\left(\begin{array}{cc}                   S_\B      &\O               \\
                                                  \O &I                \\\end{array}\right)
=S.
$$
Thus, the following result holds (the second assertion is proved as in Theorem\:\ref{250810th1}).

\begin{proposition}
\label{p_XZgen}
$\ms$\\\indent
{\rm 1.} In the presence of nonbasic agents\/$,$ the orthogonal projection onto the subspace of convergence to consensus of the DeGroot algorithm has the representation $S=XZ^{-1},$ where $X$ and $Z$ are defined in~\eqref{e_XZgen}.

{\rm 2.} The row vector $\aa^{\rm T}$ such that $\Pin=\bm1\aa^{\rm T}$ coincides with the first row of~$Z^{-1}.$
\end{proposition}

\section{THE REGULARIZED POWER LIMIT OF A STOCHASTIC MATRIX}
\label{s_regu}

The meaning of Eq.\,\eqref{e_aaZ}, Theorem\:\ref{250810th1}, and Proposition\:\ref{p_XZgen} is beyond the scope of the consensus problem.
In particular, they state that for any proper stochastic matrix $P,$ the matrix $\Pin=\Pbes S\,$ is stochastic, has rank~1 (and thus all its rows are identical) and in the case where $P$ is regular, $\Pin=\Pbes.$

Moreover, item\:3 of Theorem\:\ref{250810th1} states that the components of the vector $\aa$ that determines $\Pin$ ($\Pin=\bm1\aa^{\rm T}$) have a natural property, namely, within the same ($i$th) basic bicomponent, their ratio is equal to the ratio of the corresponding components of the vector $\pi^i,$ the stationary vector of the bicomponent. Otherwise, if two vertices belong to different bicomponents $i$ and $j$, then to obtain their ratio, the ratio of the corresponding elements of $\pi^i$ and $\pi^j$ is to be multiplied by the ratio of specific ``weights'' of the bicomponents measuring their internal homogeneity.

On the other hand, studying the problem of consensus leads to the conclusion that there are cases where it is useful to associate with a stochastic matrix $P$ an \emph{adjusted power limit of rank\/~$1$}, even though $P$ can be not regular.

These considerations allow us to call the matrix $\Pin$ associated with a proper stochastic matrix $P$ the \emph{regularized power limit
of~$P$}.
The meaningfulness of this concept is supported by the following properties.

\begin{proposition}
\label{p_regul}
For any proper stochastic matrix $P$\/{\rm:}

{\rm(1)} vector $\aa$ such that $\Pin=\bm1\aa^{\rm T}$ is a stationary vector of the matrix $P\!:$ $\aa^{\rm T}P=\aa^{\rm T};$

{\rm(2)} $\Pin P=P\hspace{-.1em}\Pin=\Pbes\!\Pin=\Pin\Pbes=\Pin.$
\end{proposition}

Item~1 of Proposition\:\ref{p_regul} follows from the fact that by item\:3 of Theorem\:\ref{250810th1}, $\aa$ is a linear combination of the vectors $\tilde\pi^1\cdc\tilde\pi^\nu$ each of which is a stationary vector of~$P.$ Item\:2 follows from item\:1.

Note that using the Ces\`aro limit $\lim_{m\to\infty}\frac1m\,\sum_{k=1}^m P^k,$ the concept of the regularized power limit of a stochastic matrix can be extended to arbitrary (not necessarily proper) stochastic matrices.

\section{CONCLUSION}

In this paper, we considered the problem of reaching consensus in the case where the influence matrix appearing in the DeGroot model is proper, but not necessarily regular. To solve this problem, we propose the orthogonal projection method. On the first stage, this method projects the space of initial opinions onto the region $T_P$ of convergence to consensus of the DeGroot algorithm; this stage is called preequalization. On the second stage, the result of the first stage is transformed into consensus by the iterative adjustment with the initial influence matrix. The subspace $T_P$ is the direct sum of $\RR(L),$ where $L=I-P,$ and the linear span of the vector $\bm1$ consisting of ones. We studied the properties of the method and obtained an interpretation of the resulting weights of agents in terms of spanning out-trees in the communication digraph. It is shown that the resulting matrix $\Pin=\Pbes S$ of the orthogonal projection procedure, where $S$ is the orthogonal projection onto $T_P$ and $\Pbes=\lim_{k\to\infty}P^k,$ can be considered as the regularized power limit of the stochastic matrix~$P.$

\appendix{}

\PTH{\ref{240810pr1}}
From $\,{L\bm1=\bm0}\,$ it follows that $\,\bm1\in\NN(L),$ and by \eqref{e_NLRL}, $\bm1\notin\RR(L).$
Therefore, the sum of $\RR(L)$ and $T_1$ is direct. Suppose that a vector of initial opinions $x$ belongs to $\RR(L)\oplus T_1$. Then $x=v+a\bm1,$ where $v\in\RR(L)$ and $a\in\R$. By \eqref{e_PLLP}, $\,\Pbes x=\Pbes(v+a\bm1)=a\bm1,$ i.e., consensus is achieved: $x\in T_P$.

Suppose now that for a vector of initial opinions $x,$ a consensus is achieved: $x\in T_P.$ Then $\Pbes x=a\bm1$ for some $a\in\R.$ Since $\Pbes\bm1=\bm1,$ we have $\Pbes (x-a\bm1)=\bm0,$ from which $x-a\bm1\in\NN(\Pbes).$ By \eqref{e_compl} $\NN(\Pbes)=\RR(L),$ therefore, $x-a\bm1\in\RR(L).$ Consequently, for some $v\in\RR(L)$ we have $x=v+a\bm1,$ and so $x\in\RR(L)\oplus T_1.$
\epr

\PCR{\ref{c_Li1}}
According to~\eqref{e_ranks}, $\rank L=n-\nu.$
In the proof of Proposition~11 in \cite{AgaChe00}, it is shown that the diagonal block of $L$ corresponding to the union of the nonbasic bicomponents of $\G$ has full rank. On the other hand, each diagonal block $L_i$ corresponding to a basic bicomponent has rank one less than its order and the sum of the columns of this block is~$\bm0.$ Therefore, constructing a maximal set of linearly independent columns of $L$ and having chosen all columns corresponding to the nonbasic bicomponents, among the columns corresponding to any basic bicomponent, one can reject no more than one, thus, exactly one column, since the number of rejected columns must be~$\nu.$ In the representation of the rejected column corresponding to the $i$th basic bicomponent by a linear combination of a maximal set of independent columns, the coefficients of the columns that correspond to the $i$th bicomponent are~$\/-1.$ 
Thus, removing from $L$ a single \emph{arbitrary\/} column for each basic bicomponent and adding all ``nonbasic'' columns, we obtain a maximal set of linearly independent columns of~$L.$ Owing to Theorem\:\ref{240810pr1}, for matrix $U$ composed of a maximal set of linearly independent columns of $L$ and the column $\bm1,$ we have $\RR(U)=T_P.$
\epr

\PTH{\ref{p_Pin}}
Since the nonbasic vertices correspond to the inessential states of the Markov chain determined by $P$, $\Pbes$ has the form $(\ast\;\O)$, where block $\O$ consists of $n-b$ columns \cite[(102) in Chapter\:13]{Gantmacher60}. According to \eqref{e_Sbl},
$S\!=\!\left(\!\!\!\begin{array}{cc}S_\B&\!\O\\
                                               \O&\!I\\\end{array}\!\!\!\right)\!,$
therefore, $\Pin=\Pbes S,$ as well as~$\Pbes,$ has the form $(\ast\;\O).$
Using \eqref{e_Pinaa} we obtain $\aa=(\aa_1\cdc\aa_b,0\cdc0)^{\rm T}$.

By virtue of \eqref{051110eqa} the upper left blocks of order $b$ of $P^2,P^3,\ldots$ and $\Pbes$ are $(P_\B)^2,(P_\B)^3,\ldots$ and $(P_\B)^\infty,$ respectively. Using \eqref{e_Sbl} we obtain that the upper left block of order $b$ of $\Pin=\Pbes S$ is $(P_\B)^\infty S_\B,$ which completes the proof of Theorem\:\ref{p_Pin}.
\epr

\PLE{\ref{041110pr1}}
We first prove that $q^2\cdc q^\nu$ are linearly independent.
Indeed, in the opposite case, $\sum_{i=2}^\nu\aa_iq^i=0,$ where some coefficients $\aa_i$ are nonzero. Let $j=\min\,\{\,i\mid\aa_i\ne0\}.$ Then by definition, $\aa_jq^j$ contains the nonzero components of $\aa_j\pi^{j-1},$ while the corresponding components of all vectors $q^i$ with $i>j$ are equal to zero. We obtain that $\sum_{i=2}^\nu\aa_iq^i\ne0,$ which proves the linear independence of $q^2\cdc q^\nu$.

Let $q^1$ be $\bm1,$ the column of $n$ ones.
Assume that $q^1,q^2\cdc q^\nu$ are dependent. Then $q^1=\sum_{k=2}^\nu\aa_kq^k$ for some $\aa_2\cdc\aa_\nu$ (as $q^2\cdc q^\nu$ are independent). Since the first $m_1$ components of $q^2$ are positive and the corresponding components of $q^i$  are equal to zero for $i>2,$ we have $\aa_2>0$. The next $m_2$ components are negative in $q^2,$ positive in $q^3,$ and equal to zero in the remaining vectors $q^i$, therefore, $\aa_3>0$. Proceeding by induction we obtain $\aa_\nu>0$. On the other hand, the last $m_\nu$ components of $q^\nu$ are negative, which contradicts $q^1=\sum_{k=2}^\nu\aa_kq^k$. Thus, $q^1\cdc q^\nu$ are linearly independent.

By definition, $q^2\cdc q^\nu\in\RR(\Pbes).$ According to \eqref{e_compl}, $\RR(\Pbes)=\NN(L).$ Hence, $q^2\cdc q^\nu\in\NN(L).$ Furthermore, $q^1\in\NN(L)$ as $L\bm1=0.$ Thus, the whole set of columns of $Z$ consists of the vectors $q^1\cdc q^\nu$ forming a linearly independent subset in $\NN(L)$ and a collection of linearly independent columns of $L$ that belong to $\RR(L).$ Finally, by \eqref{e_NLRL}, $\NN(L)\cap\RR(L)=\{\bm0\}.$ Therefore, the columns of $Z$ are independent and $Z$ is nonsingular.
\epr

\PTH{\ref{250810th1}}

1. Consider Eq.\,\eqref{e_aaZ}. It follows from the definition of $X,$ stochasticity of $\Pbes,$ and the identity $\Pbes L=0$ \eqref{e_PLLP} that the first column of $\Pbes X$ consists of ones, while the remaining columns consist of zeros. Therefore, each row of the resulting matrix $\Pbes XZ^{-1}=\bm1\aa^{\rm T}$ is equal to the first row of~$Z^{-1}$.

2. To prove this assertion, we use the following lemma.

\begin{applemma}
\label{l_l}
If all the entries in the first column of $A\in\C^{n\times n}$ are equal to $1$ and $A$ is invertible\/$,$ then $A^{-1}\bm1=(1,0\cdc0)^{\rm T}.$
\end{applemma}

\PLE{\ref{l_l}}
By $A^{ij}$ we denote the cofactor of the element $a_{ij}$ of $A.$ Let $\si_k$ be the sum of the entries in the $k$th row of $A^{-1}.$
Under the assumptions of Lemma\:\ref{l_l}, the expansion of $\det A$ along the first column yields
\begin{gather*}
\si_1
=\frac{\sum_{i=1}^nA^{i1}}{\det A}
=\frac{\sum_{i=1}^nA^{i1}}{\sum_{i=1}^n(1\cdot A^{i1})}=1.
\end{gather*}

By $M^{i|k}$ and $M^{ij|km}$ we denote the minor of $A$ obtained by removing row $i$ and column $k$ and
                                       the minor        obtained by removing rows $i$ and $j$ and columns $k$ and~$m,$ respectively.
Expanding the minors $M^{i|k}$ of order $n-1$ along the first column, for $k>1$ we obtain
\beq
\label{e_Ark}
\si_k\det A
=\sum_{i=1}^n(-1)^{i+k}M^{i|k}
=\sum_{i=1}^n(-1)^{i+k}\Bigl(\sum_{j=1}^{i-1}(-1)^{1+j}M^{ij|1k}+\sum_{j=i+1}^n(-1)^{j}M^{ij|1k}\Bigr).
\eeq
For any $i,j\in\{\1n\}$ such that $j<i,$ the minor $M^{ij|1k}$ enters the sum \eqref{e_Ark} with the sign $(-1)^{i+k+1+j},$ while the minor $M^{ji|1k}$ (having the same value) enters it with the sign $(-1)^{j+k+i}.$ Hence, these minors cancel out. Consequently, if $\det A\ne0$ and $k>1,$ then $\si_k=0$ holds. The lemma is proved.
\epr

\begin{appcorollary}[of Lemma\:\ref{l_l}]
\label{c_l}
If all the entries in the $k$th column of $A\in\C^{n\times n}$ are equal to $y\in\C$ and $A$ is invertible\/$,$ then $A^{-1}\bm1=(0\cdc0,y^{-1},0\cdc0)^{\rm T},$ where $y^{-1}$ is the $k$th component of the vector.
\end{appcorollary}

Corollary\:\ref{c_l} is obvious.

Now we prove item\:2 of Theorem\:\ref{250810th1}. The identity $\sum_{i=1}^n\aa_i=1$ was proved in the last paragraph of Section\:\ref{s_metho}. Since the first column of $Z$ consists of ones, the row sums of $Z^{-1},$ except for the sum of the first row (which is equal to~$1$), are equal to zero by Lemma\:\ref{l_l}.

It remains to prove the positivity of the elements of the first row of $Z^{-1}.$
Denote by $t^i$ and $t^i_k$ the total weight of all spanning out-trees of the $i$th bicomponent of $\G$ and the total weight of those of them that are rooted at the $k$th vertex of the $i$th bicomponent, respectively.
According to the matrix tree theorem (see, e.g., Theorem~VI.27 in \cite{Tutte84} or Theorem~16.9$'$ in \cite{Harary69}, which is formulated for the matrix $L^{\rm T}$ and unweighted digraphs) $t^i_k$ is equal to the cofactor of any element in the $k$th row of $L_i$ (see\:\eqref{191010eq1}).
Let $W_i$ be the determinant of the matrix obtained from $L_i$ by replacing the first column by the vector~$\pi^i.$ Expanding $W_i$ along the first column and using~\eqref{e_J1} we obtain
\begin{gather*}
W_i
=\sum_{k=1}^{m_i}\pi^i_k\,t^i_k
=\sum_{k=1}^{m_i}(\pi^i_k)^2\,t^i
=\sum_{k=1}^{m_i}\frac{(t^i_k)^2}{t^i}.
\end{gather*}

Find the cofactors $Z^{h1}$ ($h=\1n$) of the elements in the first column of $Z$ (see\:\eqref{e_Z}). Here, we represent the number $h$ of the row in the form $h=\sum_{u=1}^{c(h)-1}m_u+k(h),$ where $c(h)$ is the number of the basic bicomponent containing vertex~$h$ and $k(h)$ is the number of this vertex in the bicomponent with number $c(h)$.
Let $Z^{\langle h1\rangle}$ be the matrix $Z$ after deleting the $h$th row and the first column. To find
\beq
\label{e_det1}
Z^{h1}=(-1)^{h+1}\det Z^{\langle h1\rangle},
\eeq
in $Z^{\langle h1\rangle}$ we sequentially move each column $q^{u+1}$ (where $u=1\cdc c(h)-1$) $\xy m_u-1$ steps to the left (if $c(h)=1,$ then no columns are moved). Denoting the resulting determinant by $\tilde Z^{h1}$, we have
\beq
\label{e_det2}
\det Z^{\langle h1\rangle}
=(-1)^{\sum_{u=1}^{c(h)-1}(m_u-1)}\tilde Z^{h1}
=(-1)^{h-k(h)-(c(h)-1)}\tilde Z^{h1}.
\eeq
Furthermore, the determinant $\tilde Z^{h1}$ is equal to the product of its diagonal subdeterminants of orders $m_1\cdc m_{c(h)-1},m_{c(h)}-1,m_{c(h)+1}\cdc m_\nu,$ since the products containing other elements of the columns $q^u$ (including the columns that were moved) are equal to zero. Consequently,
\beq
\label{e_det3}
\tilde Z^{h1}
=\!\!\prod_{u=1}^{c(h)-1}\!\!W_u(-1)^{k(h)+1}\,t_{k(h)}^{c(h)}(-1)^{\nu-c(h)}\!\!\prod_{u=c(h)+1}^\nu\!\!W_u
=(-1)^{\nu-c(h)+k(h)+1}\,t_{k(h)}^{c(h)}\!\!\prod_{u\ne c(h)}W_u,
\eeq
where the factor $(-1)^{\nu-c(h)}$ appears due to the negativity of the first columns of the last $\nu-c(h)$ diagonal subdeterminants. Substituting \eqref{e_det3} into \eqref{e_det2} and \eqref{e_det2} into \eqref{e_det1} we obtain
\beq
\label{e_det4}
Z^{h1}=(-1)^{\nu+1}\,t_{k(h)}^{c(h)}\prod_{u\ne c(h)}W_u.
\eeq

From \eqref{e_det4} and item\:1 of Theorem\:\ref{250810th1} it follows that the signs of all components of $\aa$ are the same; note that $t_{k(h)}^{c(h)}$ are nonzero. Since, as shown in Section\:\ref{s_metho}, the sum of the components of $\aa$ is equal to $1$, all these components are positive.

Prove item\:3 of Theorem\:\ref{250810th1}.
For simplicity, we use the following notation: $i\!=\!c(g),$ ${j\!=\!c(h),}$ $k\!=\!k(g),$ and $r\!=\!k(h).$
Due to item\:1 of Theorem\:\ref{250810th1}, \eqref{e_det4} and \eqref{e_J1}, we have
\begin{eqnarray}
\label{e_aparo}
\frac{\aa_g}
     {\aa_h}
&=&\frac{Z^{g1}}
        {Z^{h1}}
  =\frac{t^i_k/W_i}
        {t^j_r/W_j}
  =\frac{t^i_k/\sum_{l=1}^{m_i}\pi^i_l\,t^i_l}
        {t^j_r/\sum_{l=1}^{m_j}\pi^j_l\,t^j_l}
  =\frac{\pi^i_k\,t^i/\sum_{l=1}^{m_i}\pi^i_l\,t^i_l}
        {\pi^j_r\,t^j/\sum_{l=1}^{m_j}\pi^j_l\,t^j_l}\\ 
&=&\frac{\pi^i_k(t^i)^2/\sum_{l=1}^{m_i}(t^i_l)^2}
        {\pi^j_r(t^j)^2/\sum_{l=1}^{m_j}(t^j_l)^2}
  =\frac{\bb_i\pi^i_k}
        {\bb_j\pi^j_r}
  =\frac{t^i_k\,t^i/\sum_{l=1}^{m_i}(t^i_l)^2}
        {t^j_r\,t^j/\sum_{l=1}^{m_j}(t^j_l)^2},\nonumber
\end{eqnarray}
from which, in particular, we obtain the desired statement. The theorem is proved.
\epr